\documentclass[a4paper,11pt]{article}
\pdfoutput=1 

\usepackage{jcappub} 

\usepackage{graphicx}
\usepackage{amsmath}
\usepackage{amssymb}
\usepackage{rotating} 
\usepackage{xspace}
\usepackage{color}
\usepackage{xcolor}
\usepackage{booktabs}
\usepackage[normalem]{ulem}
\usepackage{lscape}
\usepackage{adjustbox}
\usepackage{multirow}

\newcommand{\bfx}{{\bf x}}
\newcommand{\bfv}{{\bf v}}
\newcommand{\Msun}{{\rm M}_\odot}

\title{Velocity-dependent J-factors for Milky Way dwarf spheroidal analogues in cosmological simulations}

\author[a]{Keagan Blanchette,}
\author[b]{Erin Piccirillo,}
\author[c, a]{Nassim Bozorgnia,}
\author[b]{Louis E. Strigari,}
\author[d]{Azadeh Fattahi,}
\author[d]{Carlos S. Frenk,}
\author[e]{Julio F. Navarro,}
\author[f]{and Till Sawala}

\affiliation[a]{Department of Physics and Astronomy, York University,\\  4700 Keele Street, Toronto, Ontario M3J 1P3 Canada}
\affiliation[b]{Department of Physics and Astronomy,\\ Mitchell Institute for Fundamental Physics and Astronomy,\\Texas A\&M University, College Station, TX 77843, USA}
\affiliation[c]{Department of Physics, University of Alberta, \\
Edmonton, Alberta T6G 2E1, Canada}
\affiliation[d]{Institute for Computational Cosmology, Durham University,\\
South Road, Durham DH1 3LE, UK}
\affiliation[e]{Department of Physics and Astronomy, University of Victoria,\\ Victoria, BC V8P 5C2, Canada}
\affiliation[f]{Department of Physics, Gustaf H\"{a}llstr\"{o}min katu 2,\\ University of Helsinki, Finland}

\abstract{We study the impact of the dark matter velocity distribution modelling on signals from velocity-dependent dark matter annihilation in Milky Way dwarf spheroidal galaxies. Using the high resolution APOSTLE simulations, we identify analogues corresponding to Milky Way dwarf spheroidal galaxies, and from these directly determine the dark matter pair-wise relative velocity distribution, and compare to best-fitting Maxwell-Boltzmann distribution models. For three velocity-dependent annihilation models, p-wave, d-wave, and the Sommerfeld model, we quantify the errors introduced when using the Maxwell-Boltzmann parameterization. We extract a simple power-law relation between the maximum circular velocity of the dwarf spheroidal analogue and the peak speed of the Maxwell-Boltzmann distribution. We show that this relation can be used to accurately calculate the dark matter relative velocity distribution, and find that it allows us to estimate the dark matter annihilation signal without the need to directly calculate the relative velocity distribution for each galaxy. The scatter in the J-factors calculated from the analogues dominates the uncertainty obtained when compared to the J-factor as determined from the observational data for each dwarf spheroidal, with the largest scatter from d-wave models and the smallest from Sommerfeld models.}

\begin{document}
\maketitle
\flushbottom

\section{Introduction}
\label{sec:intro}
Indirect dark matter (DM) searches strive to identify Standard Model particles produced through the annihilation or decay of DM particles in astrophysical environments~\cite{Conrad:2015bsa,Gaskins:2016cha}. These searches require identifying environments of high DM density, and understanding how DM is distributed in these environments. Because of their high DM-to-luminous mass ratios as obtained from their stellar kinematics, and their relative proximity, dwarf spheroidal galaxies (dSphs) of the Milky Way (MW) are ideal candidates for indirect DM searches~\cite{Strigari:2018utn}, in particular for searches using high energy gamma rays. Fermi-LAT observations of gamma rays from dSphs provide the most stringent exclusion limits on the DM annihilation cross section for DM masses up to $\sim$~TeV~\cite{Fermi-LAT:2016uux}, while H.E.S.S.~\cite{H.E.S.S.:2020jez} and HAWC~\cite{Albert:2017vtb} set the strongest bounds for  DM masses greater than 1 TeV.

The aforementioned bounds on the annihilation cross section obtained from dSphs typically assume the s-wave DM annihilation model in which the annihilation cross section is velocity-independent. However, most generally, the DM annihilation cross section may depend on the relative DM velocity, in which case the phenomenology differs from that of pure s-wave models. For example, in Sommerfeld enhanced models, the annihilation is typically enhanced in dSphs relative to the s-wave models, because the cross section is inversely proportional to the velocity~\cite{Essig:2010em}. On the other hand, in the case of p and d-wave models, the cross section is proportional to the second and fourth powers of the velocity, respectively~\cite{Zhao:2017dln,Boddy:2019wfg}. In these latter two cases, the emission from dSphs is expected to be suppressed due to their low velocity dispersions. 

\par The density and velocity dependence of the DM annihilation signal is parametrized by the $\mathcal{J}$-factor. 
For annihilation cross sections that are independent of relative DM velocity, or s-wave models, the \emph{${\cal J}$-factor} is simply reduced to an integral over the square of the DM density profile.  To account for the p-wave, d-wave, or Sommerfeld models, the $\mathcal{J}$-factor must also incorporate the DM pair-wise relative velocity distribution \cite{Robertson:2009bh,Ferrer:2013cla,Boddy:2017vpe,Zhao:2017dln,Petac:2018gue,Boddy:2018ike,Boddy:2019wfg, Board:2021bwj,Piccirillo:2022qet,Lacroix:2021psm}. Such  $\mathcal{J}$-factors have been estimated for dSphs in previous work for different annihilation channels \cite{Essig:2010em,Zhao:2017dln,Boddy:2019wfg,Boddy:2019qak}.

The previous determinations of the velocity-dependent $\mathcal{J}$-factors from dSphs rely on assumptions for the form of the DM relative velocity distribution. Two common assumptions are 1) a Maxwell-Boltzmann (MB) distribution, and 2) an isotropic model in which the relative velocity distribution is derived from the density profile under an equilibrium assumption using an Eddington inversion. While both of these models may be well-motivated from theoretical considerations,  they have yet to be rigorously tested against the distributions in dSph-like subhalos in cosmological simulations. 

\par In this paper, we extract the DM density and relative velocity distributions of the MW's dSphs from the APOSTLE hydrodynamical simulations~\cite{Fattahi_2016,Sawala:2014xka}, and compute the $\mathcal{J}$-factors for the simulated dSphs for the s-wave, p-wave, d-wave, and Sommerfeld models. To extract the relative velocity distributions, we first identify analogues of MW satellite galaxies in the APOSTLE simulations by matching observed properties such as the circular velocity at the half-light radius and stellar mass to the corresponding subhalos in the simulation. From these best analogue candidates, we extract the DM relative velocity distribution, and thereby the $\mathcal{J}$-factors for the MW analogue subhalos. As a result, we are able to take a step towards estimating a quantity akin to a cosmic halo-to-halo scatter in the $\mathcal{J}$-factors, both for the case of velocity independent and velocity-dependent models. 

\par As an additional key component of our analysis, we compare the DM relative velocity distributions from the subhalos to MB distributions, from which we ascertain how well the MB distribution works over the entire range of resolved subhalo mass scale in APOSTLE. Our comparison to the MB distribution, as well as the calculation of the $\mathcal{J}$-factors from cosmological simulations, is similar in spirit to recent analyses~\cite{Board:2021bwj} of the $\mathcal{J}$-factor and the expected signal from the MW's smooth halo in velocity-dependent DM annihilation cross section model using the Auriga~\cite{Grand:2016mgo} and  APOSTLE~\cite{Fattahi_2016,Sawala:2014xka} hydrodynamical simulations. These results find that the DM relative velocity distribution is well-described by the
MB distribution, in particular in simulations with the full effects of baryonic physics included.

This paper is organized as follows. In section \ref{sec:sim} we discuss the details of the simulations that we use. In section~\ref{sec:dSphs}, we discuss the properties of the dSph analogues. We discuss our criteria for selecting them in section~\ref{sec:dSphs-selection}, and compute their density profiles and relative velocity distributions in sections \ref{sec:dSphs-DM-dens} and \ref{sec:f-v}, respectively. In section~\ref{sec:J-fac} we outline the calculation of the $\mathcal{J}$-factors for the different DM annihilation models we consider for a variety of dSphs. In section~\ref{sec:results} we present the results and compare to those previously found in the literature. Finally, in section~\ref{sec:disc} we discuss our results and summarize our conclusions.

\section{Simulations}
\label{sec:sim}

In this work we use a set of hydrodynamical simulations of MW-mass halos from the APOSTLE project~\cite{Fattahi_2016,Sawala:2014xka}. APOSTLE simulations use the same code as the EAGLE project \cite{Schaye:2014tpa, Crain_2015} with the EAGLE reference model Ref-L100N1504 calibration, applied to zoom simulations of the Local Group analogue systems containing two MW-mass halos. The EAGLE simulations use a modified version of the \textsc{Gadget-3} Tree SPH code \cite{Springel:2008cc}, the \textsc{anarchy} version of SPH \cite{Schaye:2014tpa,Schaller:2015vsa}, and a galaxy formation subgrid model that includes photoionization, metal-line cooling, star formation, and feedback from active galactic nuclei and star formation. The cosmological parameters are from WMAP-7: $\Omega_m = 0.272$, $\Omega_b = 0.0455$, $h = 0.704$. We use five APOSTLE volumes simulated at the highest resolution available, which we refer to AP-L1 (i.e.~Level 1 or high resolution). In each of these five volumes, we choose one MW-like halo and select all of the self-bound substructures (subhalos) within the virial radius\footnote{Virial quantities are defined as those corresponding to a spherical radius where the mean enclosed matter density is 200 times the critical density.} of the MW-like halo for further study. The DM particle mass at this resolution is $m_{\rm DM} \simeq 5 \times 10^4~\Msun$, the initial gas particle mass is $m_g \simeq 1.0 \times 10^4~\Msun$, and the maximum physical softening length is $\epsilon = 134$~pc.

For the analysis in this work, the stars and DM particles are identified as belonging to a subhalo using the \textsc{subfind} algorithm \cite{Springel:2000qu}. Here a subhalo is specified with a subset of particles within the main halo which are self-bound. We restrict our selection to include only those subhalos within 300 kpc of either the MW-like halo or the M31-like halo identified in each of the five APOSTLE volumes. We further restrict our selection to include only subhalos with nonzero stellar mass and having a total mass of $\approx [10^7 - 10^{10}]~\Msun$, which is representative of the mass range of our dSph candidates. In total this leaves us with 2074 subhalos across the ten main halos. 

\section{Dwarf spheroidal galaxy analogues}
\label{sec:dSphs}
In this section, we discuss the properties of the dSph analogues that we identify in our simulations. We begin by defining a broad matching criteria to map dSphs onto subhalos in the APOSTLE simulations, and then move onto characterizing the density profiles of these systems, and finally determine the DM velocity distributions in the analogues. 

\subsection{Selection of dSph analogues}
\label{sec:dSphs-selection}
\par The selection of specific dSph analogues was performed using two matching criteria. The first criterion involves matching the observed circular velocity at the half-light radius of the dSphs \cite{Sawala:2015cdf}. For each subhalo, we first calculate the circular velocity, $V_{c}(r) = \sqrt{GM(<r)/r}$, where $M(<r)$ is the total mass enclosed within a sphere of radius $r$ centered on the subhalo. We then compute $V_{c}(r_{1/2})$, or $V_{1/2}$, where $r_{1/2}$ is the 3D half-light radius for each of our dSph counterparts \cite{Wolf_2010}. Most generally, we require that our dSph analogues have a circular velocity at the half-light radius within 2$\sigma$ of the observed value. The only exception is for the Draco analogues, which, as described below, we require it to be within 3$\sigma$ of the observed value, due to the difficulty in identifying a matching analogue. 

\par As our second criterion, we require that the subhalos have a stellar mass that is consistent with the measured stellar mass of its observed counterpart \cite{Sawala:2015cdf}. Further, considering that typical stellar mass-to-light ratios for dSph stellar populations are in the range $\sim 1-3$, and extending this range by 50\% to increase the number of matching analogues, we take the range of dSph stellar masses that we consider to be within the range $0.5-4.5$ of the measured stellar mass of the dSph. 

\par Given the relatively small sample of satellites in our simulations, and the precise measurements of the stellar mass and circular velocity, it is expected that our matching criteria will not produce exact dSph analogues. This is particularly true when considering the distance to the dSph. In the cases in which we find a matching circular velocity and stellar mass, but there is a significant difference between the distance to the observed satellite and the distance to the simulated satellite, we simply shift the simulated satellite to a distance corresponding to that of the observed satellite~\cite{McConnachie2012}. We choose this approach because we are most interested in the $\mathcal{J}$-factors below, which are a sensitive function of the dSph distance.

\par With our criteria we identify 126 unique subhalos in AP-L1 as dSph analogues. The results of our search for analogues are shown in table~\ref{tab:dSph-veldisp-no-mass-cond}. For each dSph, the number of subhalos identified as analogues, $N$, is given in the second column of this table. Note that $N$ may contain subhalos that are analogues of multiple dSphs. For each dSph, we also show the top two best matching analogues based on their s-wave $\mathcal{J}$-factor being closest to those of ref.~\cite{Boddy:2019qak} (also shown in figure \ref{fig:Boddy-compare} in section \ref{sec:results}), as well as several properties for each analogue. In several instances, for example with Sculptor, Carina, and Sextans, we find good matches between a simulated satellite and the observed system. On the other hand, as alluded to above, for Draco we are unable to locate reasonable analogues using the matching criterion of 2$\sigma$ uncertainty for $V_{1/2}$. This is similar to what has been found in previous similar studies~\cite{Sawala2015}, as Draco is less dense than is predicted given its best matching analogues in simulations. Allowing up to 3$\sigma$ uncertainty for $V_{1/2}$, we do obtain four Draco analogues across the ten simulated halos.

\begin{table}
\small
\makebox[\linewidth]{
\begin{tabular}{|l|c|c|c|c|c|c|c|}
\hline
\multirow{2}{*}{dSph Analogue} & \multirow{2}{*}{$N$} & \multirow{2}{*}{$M_{\star}^{\rm obs}~[\rm{M}_\odot]$} & \multirow{2}{*}{$M_\star~[\rm{M}_\odot]$} & {$V_{1/2}^{\rm obs}$} & {$V_{1/2}$} & $V_{\rm max}$ &$\rm{log}_{10}(\widetilde{\cal J}_s)$ \, \\
 &  &  &  & [km$/$s] & [km$/$s] & [km$/$s] & $[\rm{GeV}^2\,\rm{cm}^{-5}]$ \\ \hline
Canes Venatici I (1) & \multirow{2}{*}{39} & \multirow{2}{*}{$2.3 \times 10^5$} & $5.66 \times 10^5$ & \multirow{2}{*}{13.2} & 14.56 & 15.39 & 17.43 \\ 
Canes Venatici I (2) &  &  & $3.45 \times 10^5$ &  & 14.53 & 16.07 &  17.44 \\ \hline
Carina (1) & \multirow{2}{*}{23} & \multirow{2}{*}{$4.3 \times 10^5$} & $2.38 \times 10^5$ & \multirow{2}{*}{11.1} & 11.30 & 13.14 &  18.52\\
Carina (2) &  &  & $1.61 \times 10^6$ &  & 11.78 & 16.87 &  18.21 \\ \hline
Draco (1) & \multirow{2}{*}{4} & \multirow{2}{*}{$2.2 \times 10^5$} & $8.91 \times 10^5$ & \multirow{2}{*}{17.5} & 14.92 & 24.32 &  18.81 \\
Draco (2) &  &  & $4.70 \times 10^5$ &  & 15.24 & 24.35 &  18.77\\ \hline
Fornax (1) & \multirow{2}{*}{4} & \multirow{2}{*}{$1.7 \times 10^7$} & $1.36 \times 10^7$ & \multirow{2}{*}{18.5} & 18.79 & 20.38 &  18.01 \\
Fornax (2) &  &  & $1.20 \times 10^7$ &  & 18.36 & 21.96 &  17.87 \\ \hline
Leo I (1) & \multirow{2}{*}{20} & \multirow{2}{*}{$5.0 \times 10^6$} & $3.27 \times 10^6$ & \multirow{2}{*}{15.6} & 15.24 & 20.37 &  17.63 \\
Leo I (2) &  &  & $3.52 \times 10^6$ & & 15.15 & 24.81 &  17.64\\ \hline
Leo II (1) & \multirow{2}{*}{52} & \multirow{2}{*}{$7.8 \times 10^5$} & $1.45 \times 10^6$ & \multirow{2}{*}{11.4} & 12.15 & 20.13 &  17.66 \\ 
Leo II (2) &  &  & $4.97 \times 10^5$ & & 12.31 & 22.01 &  17.65 \\ \hline
Sculptor (1) & \multirow{2}{*}{11} & \multirow{2}{*}{$2.5 \times 10^6$} & $6.10 \times 10^6$ & \multirow{2}{*}{15.6} &  15.73  & 27.27 & 18.56 \\
Sculptor (2) &  &  & $3.27 \times 10^6$ &  &  15.04 & 20.37 &  18.54 \\
\hline
Sextans (1) & \multirow{2}{*}{7} & \multirow{2}{*}{$5.9 \times 10^5$} & $4.58 \times 10^5$ & \multirow{2}{*}{12.3} & 12.85  & 13.02 & 17.70 \\
Sextans (2) &  &  & $1.58 \times 10^6$ &  & 12.91 & 12.98 &  17.87 \\
\hline
Ursa Minor (1) & \multirow{2}{*}{24} & \multirow{2}{*}{$3.9 \times 10^5$} & $9.30 \times 10^5$ &  \multirow{2}{*}{19.9} & 19.37 & 24.41 & 18.75  \\
Ursa Minor (2) &  &  & $7.79 \times 10^5$ &  & 18.41 & 25.97 & 18.75 \\ \hline
\end{tabular}
}
\caption{The number of subhalos identified as dSph analogues, $N$, stellar mass, $M_\star$, the circular velocity at the half-light radius, $V_{1/2}$, the maximum circular velocity, $V_{\rm max}$, and the s-wave $\mathcal{J}$-factors of our selected dSph analogues in AP-L1. The observed stellar mass, $M_\star^{\rm obs}$, and the observed circular velocity at the half-light radius, $V_{1/2}^{\rm obs}$ of the dSphs are also given in the table. Note that a given subhalo may be identified as being more than one dSph analogue.}
\label{tab:dSph-veldisp-no-mass-cond}
\end{table}

\subsection{Dark matter density profiles}
\label{sec:dSphs-DM-dens}
We now move on to characterizing the DM density profiles of the subhalos. Characterizing the density profiles are important since they enter into the calculation of the DM annihilation rate through the $\mathcal{J}$-factor. We  follow the typical assumption that the particle distributions in the subhalos are spherically symmetric, which has been shown to be a good assumption for simulated dwarf galaxies in APOSTLE~\cite{Genina:2019job}. 

\begin{figure}
\begin{centering}
\includegraphics[width=0.7\textwidth]{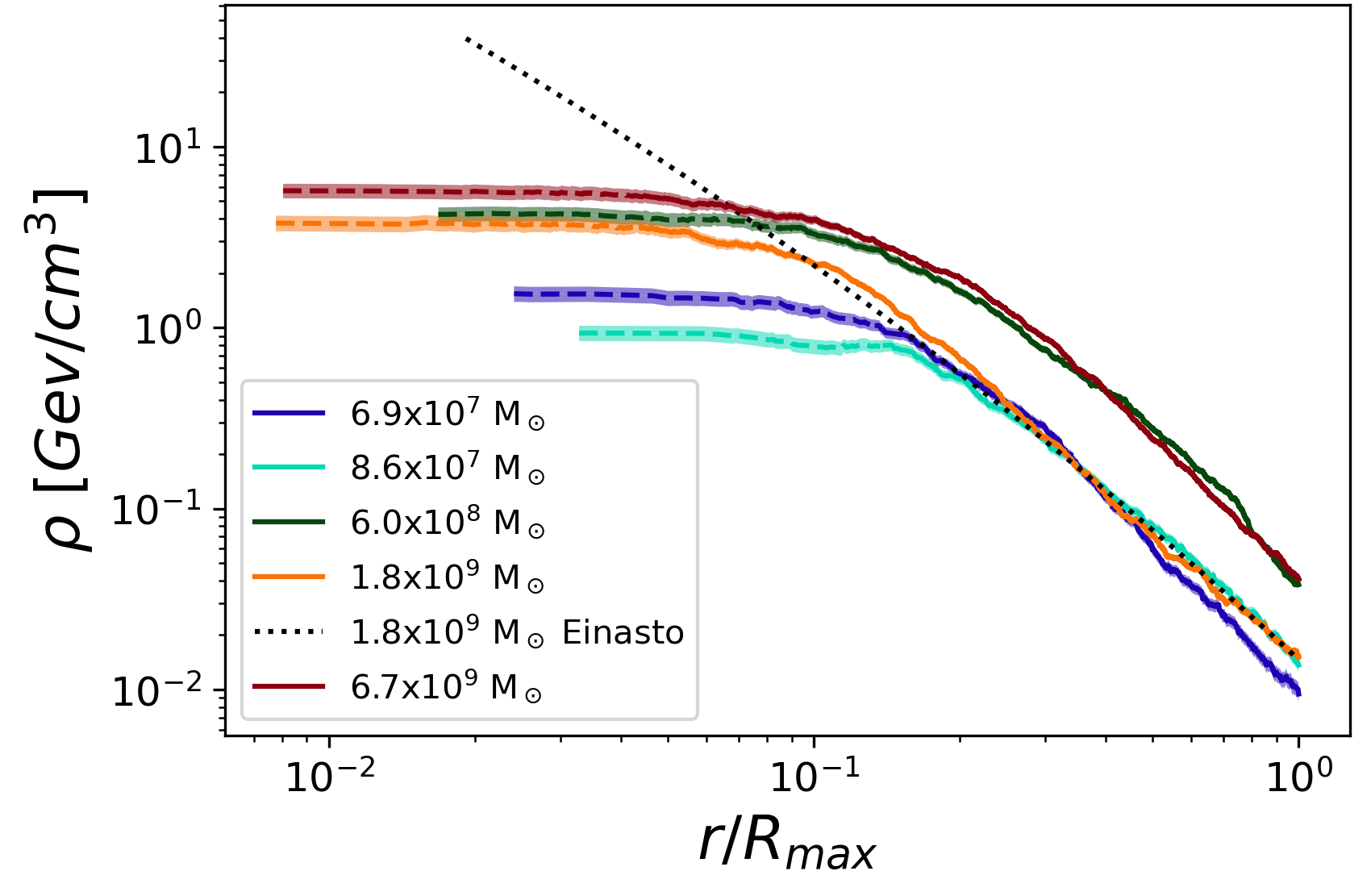}
\par\end{centering}
\caption{Spherically-averaged DM density profiles for a subset of five subhalos in the AP-L1 simulations, with masses in the range of $7\times [10^7 - 10^9]~\Msun$. The dashed portion of each curve represents the density at radii smaller than 268 pc, i.e.~twice the gravitational softening length. The lighter shaded region represents the $1 \sigma$ error, found using Poisson statistics. The dotted black line shows the best fit Einasto profile for the $1.8 \times 10^9~\Msun$ subhalo.
\label{fig:Einasto-empircal-dens}}
\end{figure}

In figure~\ref{fig:Einasto-empircal-dens} we show the spherically-averaged DM density profiles of five example subhalos
of different masses. For each simulated subhalo, the DM density is obtained from the DM mass in spherical shells of width ranging from $[0.1 - 1.0]$~kpc, and plotted as a function of $r/R_{\rm max}$, where $r$ is the distance from the subhalo center, which is identified as the center of potential of that subhalo, and $R_{\rm max}$ is the radius at which the rotation curve of the subhalo reaches its maximum value. 
The bin width of the spherical shells was chosen such that there is a minimum of 10 particles per shell.

As is shown in figure~\ref{fig:Einasto-empircal-dens}, the density profiles start to flatten towards the inner regions of the subhalos. This is due to the resolution limit of the simulations, which is determined by the gravitational softening length, $\epsilon=134$~pc. In figure~\ref{fig:Einasto-empircal-dens}, the dashed portion of the curves represent the density profiles at radii less than twice the softening length. 

\par We model the DM density profile of the simulated subhalos 
by the Einasto profile,
\begin{equation}
\rho = \rho_{-2}  \exp\left(-\frac{2}{\alpha}\left[\left(\frac{r}{r_{-2}}\right)^\alpha - 1\right]\right),
\label{eq:Einasto-profile}
\end{equation}
where $\rho_{-2}$ and $r_{-2}$ are the density and radius at which $\rho(r) \propto r^{-2}$, and $\alpha$ is a parameter which specifies the curvature of the density profile. We set this parameter to $\alpha = 0.16$~\cite{Gao:2007gh}, so only the two parameters $\rho_{-2}$ and $r_{-2}$ are varied. For each subhalo, we find the best fit Einasto profile in the range of $2 \epsilon < r < 2 R_{\rm max}$, using $2\epsilon$ rather than $\epsilon$ to be conservative and avoid resolution issues. As an example, in figure~\ref{fig:Einasto-empircal-dens} the dotted black line shows the best fit Einasto profile for the $1.8 \times 10^9~\Msun$ subhalo. It is clear that the central region of the simulated subhalos 
are under-dense compared to what is expected from the Einasto fit, which is typically true throughout all the simulated subhalos examined.

\subsection{Relative velocity distributions}
\label{sec:f-v}

\par For velocity-dependent annihilation models, the $\mathcal{J}$-factors depend not only on the DM density profile, but also on the  DM pair-wise, or relative velocity distribution in the subhalo. We now describe how we extract the DM relative velocity distributions in radial shells in each subhalo.  

\par We first extract the position vector, ${\bf x}$, and the velocity vector, ${\bf v}$, of the simulation particles belonging to each subhalo,
with respect to the center of that subhalo. Following the notation used in ref.~\citep{Board:2021bwj}, we define $f(\textbf{x}, \textbf{v})$ such that $f(\textbf{x}, \textbf{v})~d^3\textbf{x}~d^3\textbf{v}$ is the number of DM particles within a phase space volume $\textbf{x}+d^3\textbf{x}$ and $\textbf{v}+d^3\textbf{v}$. The probability distribution of DM velocities at a position ${\bf x}$ can be written as
\begin{equation}
P_\textbf{x} (\textbf{v}) = \frac{f(\textbf{x},\textbf{v})}{\rho(\textbf{x})},
\label{eq:DM-vel-prob}
\end{equation}
where the DM density at $\textbf{x}$ is given by
\begin{equation}
\rho(\textbf{x}) = \int f(\textbf{x},\textbf{v})d^3\textbf{v}.
\label{eq:DM-density}
\end{equation}

For a given pair of DM particles with velocities ${\bf v}_1$ and ${\bf v}_2$, we can write the individual velocities in terms of the center-of-mass velocity, $\textbf{v}_{\rm cm}$, and the relative velocity, $\textbf{v}_{\rm rel} \equiv \textbf{v}_2 - \textbf{v}_1$, as $\textbf{v}_1 = \textbf{v}_{\rm cm} + \textbf{v}_{\rm rel}/2$ and $\textbf{v}_2 = \textbf{v}_{\rm cm} - \textbf{v}_{\rm rel}/2$. 
We can then write a general expression for the distribution of relative velocities at a position $\textbf{x}$,
\begin{equation}
P_\textbf{x}(\textbf{v}_{\rm rel}) = \int P_\textbf{x}(\textbf{v}_1={\bf v}_{\rm cm}+{\bf v}_{\rm rel}/2) P_\textbf{x}(\textbf{v}_2 = {\bf v}_{\rm cm}-{\bf v}_{\rm rel}/2)~d^3\textbf{v}_{\rm cm}.
\label{eq:P-x-vrel}
\end{equation}

The DM relative velocity modulus distribution, $P_\textbf{x}(|\textbf{v}_{\rm rel}|)$, is related to the relative velocity distribution, $P_\textbf{x}(\textbf{v}_{\rm rel})$ by
\begin{equation}
P_\textbf{x}(|\textbf{v}_{\rm rel}|) = v^2_{\rm rel}\int P_\textbf{x}(\textbf{v}_{\rm rel})~d\Omega_{\textbf{v}_{\rm rel}},
\label{eq:P-x-modulus}
\end{equation}
where $d\Omega_{\textbf{v}_{\rm rel}}$ is an infinitesimal solid angle along the direction $\textbf{v}_{\rm rel}$. It is normalized to unity, such that $\int P_\textbf{x}(|\textbf{v}_{\rm rel}|)dv_{\rm rel} = 1$.

\par To extract the relative velocity modulus distributions for each subhalo, we define spherical shells of  fixed width progressing radially outward from the subhalo center. In each shell, we extract the three components of the velocity vectors, and find the modulus of the pairwise relative velocity distributions for all DM particles in the shell. 
Each spherical shell has a radial width of 0.2 kpc, chosen such that there are at least 10 particles in each shell.

\par
\begin{figure}
\begin{centering}
\includegraphics[scale=0.115]{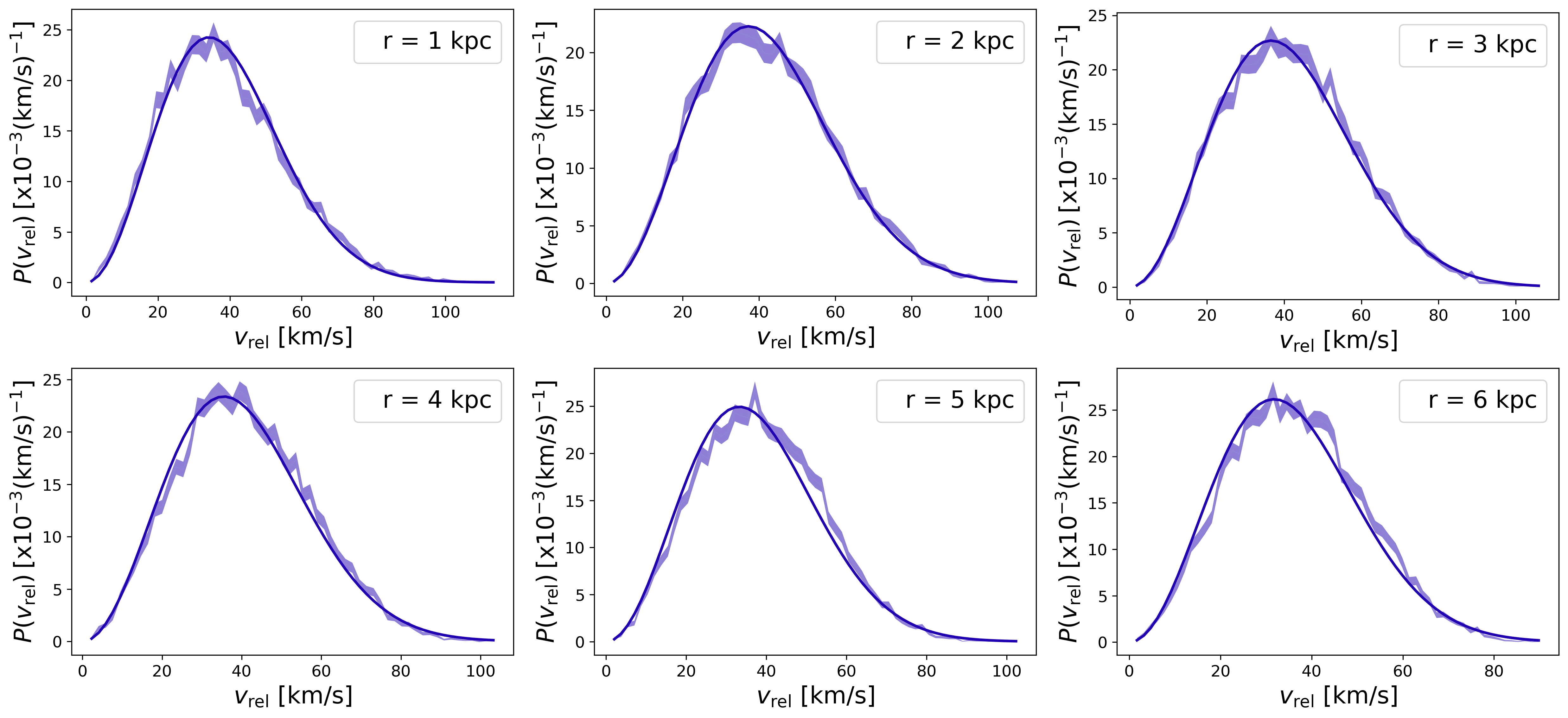}
\par\end{centering}
\caption{The DM relative velocity modulus distribution in 1 kpc shells for a typical subhalo
of mass $7.0 \times 10^9$~${\rm M}_\odot$. The different panels show six different 1 kpc shells, starting from the shell closest to the subhalo center in the upper-left, and continuing to the shell furthest from the center in the bottom-right. The purple shaded bands specify the $1\sigma$ Poisson error in the speed distributions, while the black solid lines show the best fit MB distribution in each case. \label{fig:fv}}
\end{figure}

 Figure \ref{fig:fv} shows the DM relative velocity modulus distributions for an example subhalo. The speed distributions are shown in 1 kpc spherical shells\footnote{We use spherical shells of 1 kpc width in figure \ref{fig:fv}  for clarity of presentation, while in the analysis of the J-factors we use a shell width of 0.2 kpc.} at different radii from the center of the subhalo, starting from a distance of 1 kpc from the subhalo center, and going up to 6 kpc from the center. The purple shaded bands specify the 1$\sigma$ Poisson error in the speed distributions.

Next, we compare the DM relative speed distributions with a MB distribution. In the Standard Halo Model \cite{Drukier:1986tm}, the DM velocity distribution is an isotropic MB  distribution with a most probable speed of $\sqrt{2}\sigma$, where $\sigma$ is the one dimensional velocity dispersion. In this model, the relative velocity distribution $P_\textbf{x}(\textbf{v}_{\rm rel})$ is also a MB distribution, but with a one dimensional relative velocity dispersion of $\sqrt{2}\sigma$ \cite{Ferrer:2013cla}. For each subhalo, we find the best-fit MB relative speed distribution,
\begin{equation}
P_{\rm MB}(|\textbf{v}_{\rm rel}|) = \frac{4v_{\rm rel}^2}{\sqrt{\pi}v_p^3}~\exp\left(-\frac{v_{\rm rel}^2}{v_p^2}\right),
\label{eq:MaxBdist}
\end{equation}
where $v_p$ is the best fit peak speed, i.e.~the most probable speed of the DM particles.

\par Once we have the empirical $P_{\bf x}(|\textbf{v}_{\rm rel}|)$ for each spherical shell in a subhalo, we can find the best fit peak speed, $v_p$, in eq.~\eqref{eq:MaxBdist} for each subhalo, by fitting the DM relative speed distributions in each shell to the MB distribution. In figure \ref{fig:fv} we show the best fit MB speed distribution for each of the six 1 kpc shells as solid black lines. As it is clear from the figure, the MB distribution provides a good fit to the DM relative speed distribution of the simulated subhalo at all radii. Notice that the data shown in figure \ref{fig:fv} is for a ``typical'' subhalo analogue 
and is representative of the  DM relative speed distributions of the other simulated subhalos studied in this work. 

\par Another method to determine the best fit peak speed of the MB distribution for each subhalo is to fit the relative speed distribution of all DM particles in the subhalo, instead of dividing it by shells. This process is much more computationally intensive, but provides an excellent check when compared to the mean peak speed found from the results of dividing the subhalo into different shells. 
Both of these methods lead to a power law relation between the best fit peak speed of the MB distribution and the maximum circular velocity, $V_{\rm max}$, of the subhalo, 
\begin{equation}
v_p = (1.0573 \pm 0.0157)\left(V_{\rm max}\right)^{1.0517 \pm 0.0038},
\label{eq:pow-law}
\end{equation}
where $v_p$ and $V_{\rm max}$ are in units of km\,s$^{-1}$, and the errors here represent the $1\sigma$ error on each fit parameter returned by the \textsc{lmfit} package for Python.

\par
\begin{figure}
\begin{centering}
\includegraphics[width=0.8\textwidth]{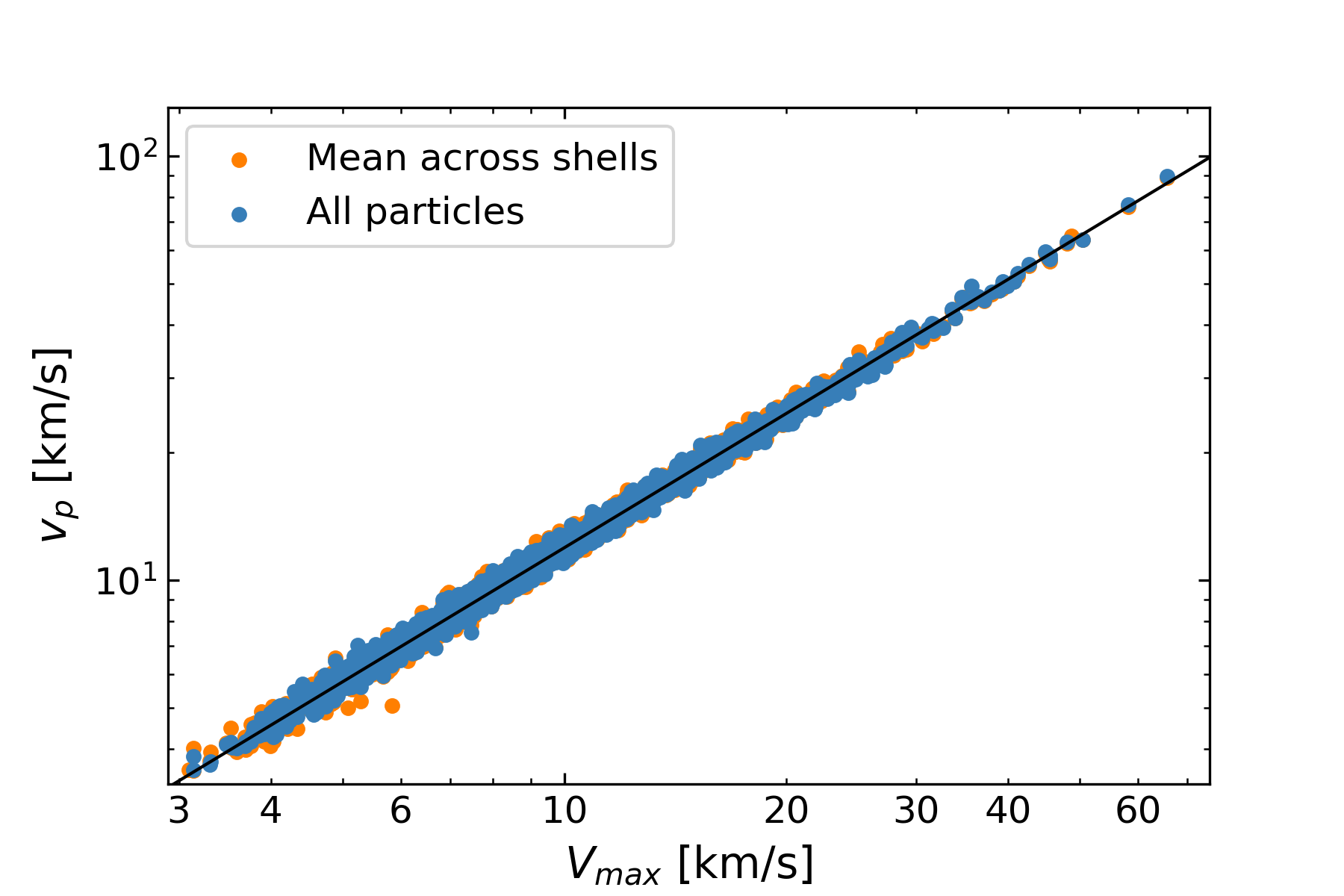}
\par\end{centering}
\caption{The best fit peak speed, $v_p$, of the MB distribution as  a function of the maximum circular velocity, $V_{\rm max}$, for all selected subhalos in AP-L1 determined by finding the best fit peak speed in different radial shells and taking the mean across all shells (orange dots), or by using a single MB fit across all particles in a subhalo (blue dots). The best fit power law (eq.~\ref{eq:pow-law}) using the orange points is shown as a solid black line.}
\label{fig:peak_v_mass_allPart-err}
\end{figure}

 Figure \ref{fig:peak_v_mass_allPart-err} shows the relation between the best fit peak speed of the MB distribution and the maximum circular velocity
of the subhalos in AP-L1 using the two methods for determining $v_p$. The blue points are computed using all the DM particles in the subhalos, and the orange points are computed by first finding the best fit peak speed for different shells, and then taking the mean of the peak speeds across all shells. Also shown in the plot is the best fit
power law using the mean of the peak speed shown as a solid line, and quantified in eq.~\eqref{eq:pow-law}. The two methods of obtaining the best fit peak speed agree within their $1\sigma$ errors.

\section{J-factors}
\label{sec:J-fac}
With the DM density profiles and relative velocity distributions now determined, we can move on to calculating the velocity-dependent $\mathcal{J}$-factors. Here we lay out the formalism for the $\mathcal{J}$-factor calculation for each of the annihilation cross section models that we consider. The notation closely follows ref.~\cite{Board:2021bwj}.

The DM annihilation cross section, $\sigma_A$, averaged over the relative velocity distribution at a spatial location, $\textbf{x}$,  is given by
\begin{equation}
\langle \sigma_A v_{\rm rel} \rangle (\textbf{x}) = \int d^3 \textbf{v}_{\rm rel} P_\textbf{x}(\textbf{v}_{\rm rel})(\sigma_A v_{\rm rel}).
\label{eq:annihilation-cross-section}
\end{equation}
In the usual s-wave annihilation, $\sigma_A v_{\rm rel}$ is independent of the relative velocity. However, for velocity-dependent annihilation models, $\sigma_A v_{\rm rel}$ depends on the relative velocity and can be parametrized as $\sigma_A v_{\rm rel} = (\sigma_A v_{\rm rel})_0 (v_{\rm rel}/c)^n$. Here $(\sigma_A v_{\rm rel})_0$ is the velocity-independent component of the annihilation cross section, and $n$ depends on the specific DM annihilation model. We consider the following cases: $n=0$ (s-wave annihilation), $n=2$ (p-wave annihilation), $n=4$ (d-wave annihilation), and $n=-1$ (Sommerfeld-enhanced annihilation).

For the general velocity-dependent annihilation, the expected gamma-ray flux from DM annihilation can then be written as
\begin{equation}
\frac{d \Phi_\gamma}{dE} = \frac{\left(\sigma_A v_{\rm rel}\right)_0}{8 \pi m_\chi^2}\frac{dN_\gamma}{dE} \mathcal{J}_s,
\label{eq:flux}
\end{equation}
where $m_\chi$ is the DM particle mass, $dN_\gamma/dE$ is the gamma-ray energy spectrum produced per annihilation, and $\mathcal{J}_s$ is the effective $\mathcal{J}$-factor defined as~\cite{Boddy:2019wfg, Board:2021bwj},
\begin{align} 
{\cal J}_s(\theta) &= \int d \ell \, \frac{\langle \sigma_A v_{\rm rel} \rangle}{(\sigma_A v_{\rm rel})_0}  \left[\rho (r(\ell, \theta))\right]^2\nonumber\\
&= \int d \ell \int d^3 \bfv_{\rm rel} P_{\bfx} (\bfv_{\rm rel}) ~\left(\frac{{v}_{\rm rel}}{c}\right)^n~ \left[\rho (r(\ell, \theta))\right]^2.
\label{eq:Jfactor}
\end{align} 
Here $\ell$ is the distance from the Sun to a point in the dSph (i.e.~line of sight), $\theta$ is the opening angle between the line of sight $\ell$ and the distance $D$ from the Sun to the center of the dSph, and $r^2(\ell, \theta) = \ell^2 + D^2 - 2\ell D \cos{\theta}$ is the square of the radial distance measured from the center of the dSph. This is with the assumption that the dSph is spherically symmetric. The $\mathcal{J}_s$-factor integrated over solid angle is then given by
\begin{equation}
\widetilde{\cal J}_s(\theta)= 2\pi \int_0^\theta {\cal J}_s(\theta')\sin\theta' d\theta'.
\end{equation}

\section{Results} 
\label{sec:results}

In this section we present the $\widetilde{\cal J}_s$-factors of the dSph analogues in AP-L1 for the different velocity-dependent annihilation models. We also quantify the  errors introduced in the $\widetilde{\cal J}_s$-factors if we model the DM relative velocity distribution of the dSph as a MB distribution.

\begin{figure}
\begin{centering}
\includegraphics[width=0.98\textwidth]{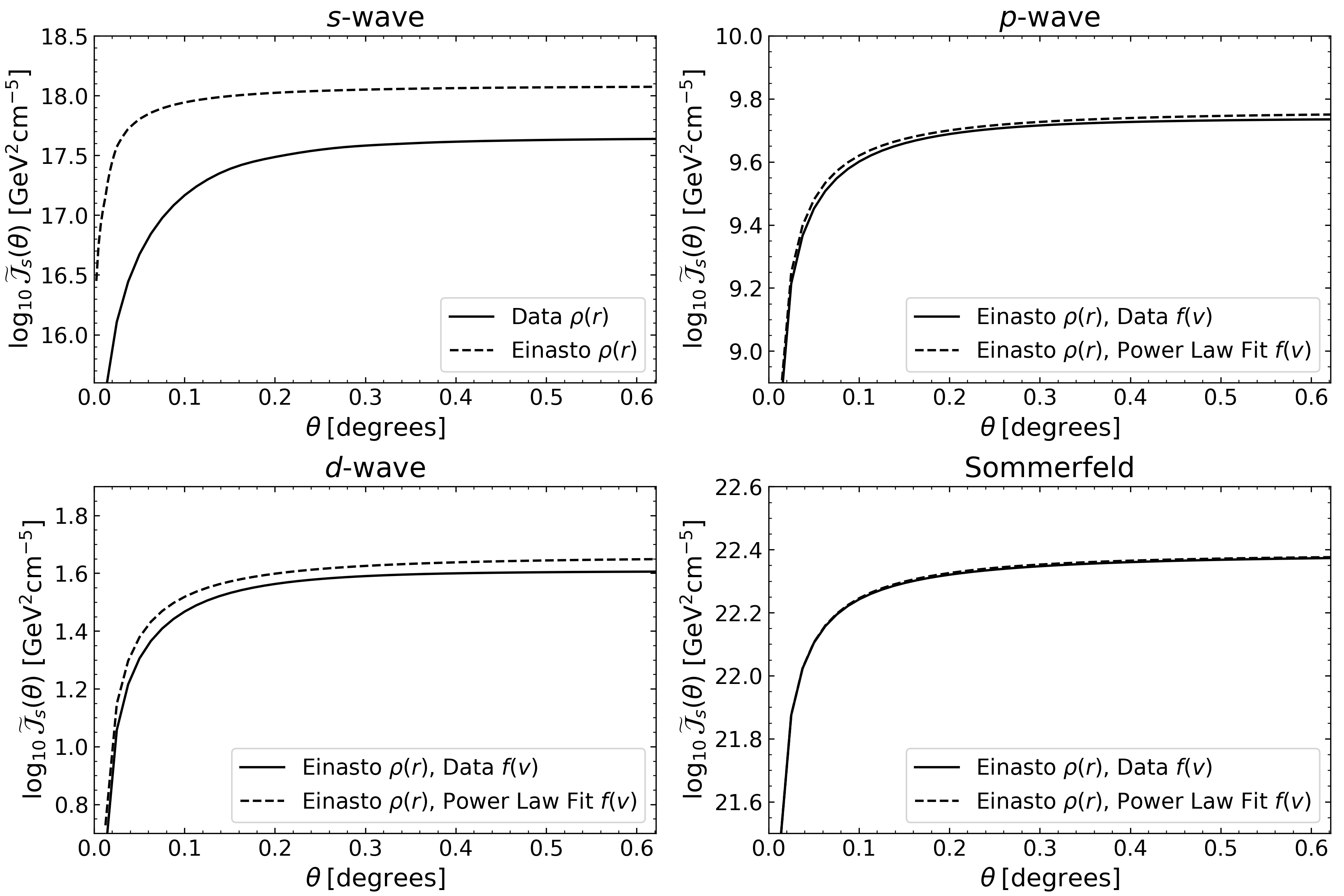}
\par\end{centering}
\caption{$\widetilde{\cal J}_s$-factor for one Carina dSph analogue in AP-L1 for the s-wave (top left), p-wave (top right), d-wave (bottom left), and Sommerfeld (bottom right) annihilation models. In the s-wave panel, the solid and dashed lines represent using the empirical and an Einasto fit to the DM density profile, respectively, for the $\widetilde{\cal J}_s$-factor calculation. In the p-wave, d-wave, and Sommerfeld panels, the solid and dashed lines represent using the empirical DM relative velocity distribution and a MB distribution with a peak speed determined from the power law relation in eq.~\eqref{eq:pow-law}, respectively.}
\label{fig:dSph-carina-J-factor-composite}
\end{figure}

\begin{figure}
\begin{centering}
\includegraphics[width=0.98\textwidth]{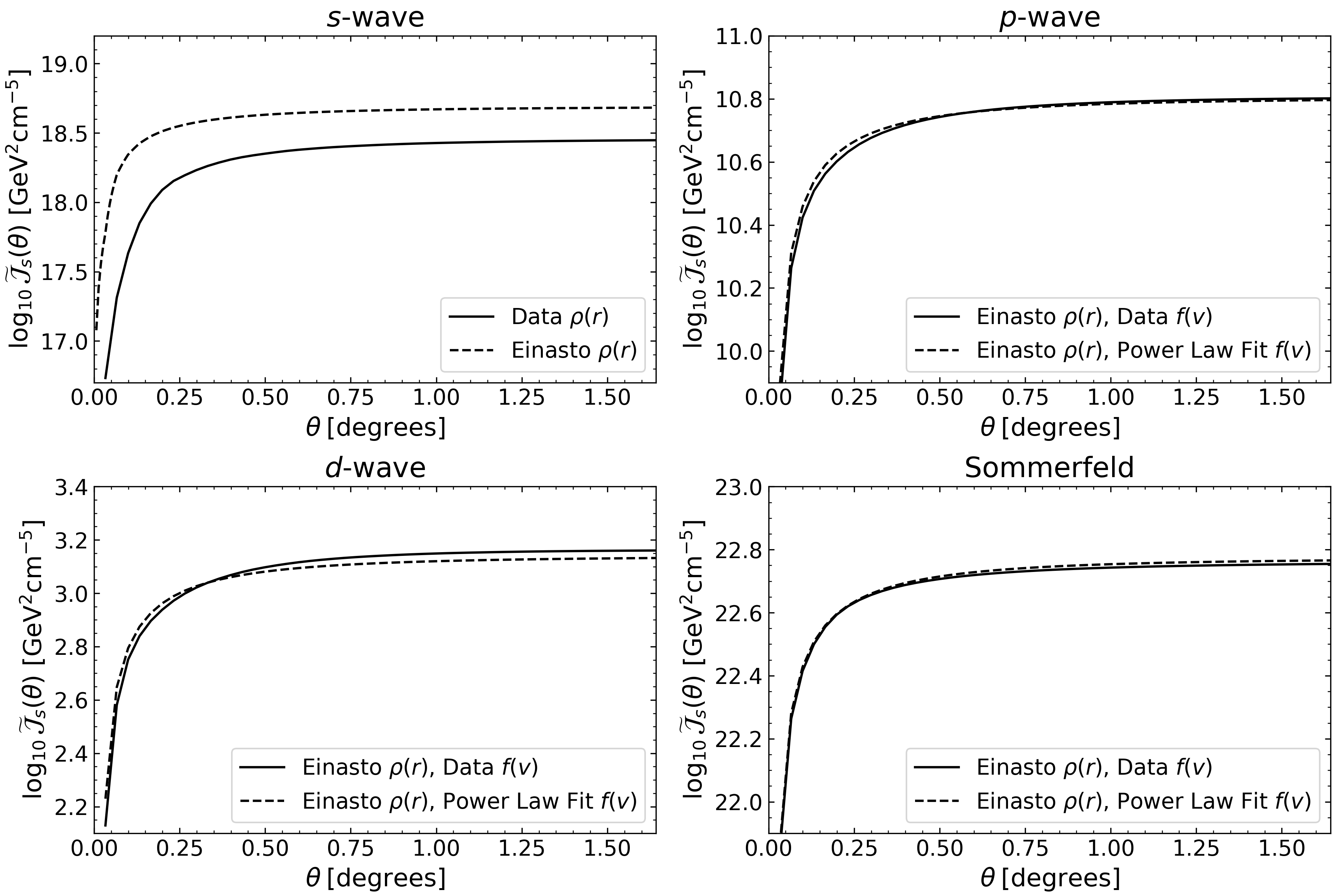}
\par\end{centering}
\caption{Same as figure \ref{fig:dSph-carina-J-factor-composite} but for one Sculptor dSph analogue in AP-L1. 
}
\label{fig:dSph-sculptor-J-factor-composite}
\end{figure}

In figures \ref{fig:dSph-carina-J-factor-composite} and \ref{fig:dSph-sculptor-J-factor-composite} we show the $\widetilde{\cal J}_s$-factors as a function of the opening angle, $\theta$, for a subset of simulated dSphs selected to be analogues of Carina and Sculptor, based on the criteria discussed in section \ref{sec:dSphs-selection}. The four panels of the figures show the $\widetilde{\cal J}_s$-factors for the four annihilation models. In the s-wave panel, the results are shown using two methods of computing the DM density profiles. In one method the density profiles are directly computed from the simulation data, and in the other method an Einasto fit to the density profiles is used. In the p-wave, d-wave, and Sommerfeld panels, the best fit Einasto density profiles are used, while two methods are employed to compute the DM relative velocity distributions. In one method the velocity distributions are extracted from the simulation data directly, and in the other method the power law relation in eq.~\eqref{eq:pow-law} is used to find the MB peak speed for each dSph analogue from its maximum circular velocity.

 We can  clearly see from the s-wave panel of figures \ref{fig:dSph-carina-J-factor-composite} and \ref{fig:dSph-sculptor-J-factor-composite} that the $\widetilde{\cal J}_s$-factors are larger when the best fit Einasto density profile is used. This is due to the empirical density profiles being under-dense in the inner regions of the dSph analogues, which results in smaller $\widetilde{\cal J}_s$-factors obtained from the simulation data compared to those obtained from the Einasto profile. It is also clear from the other panels of the figures that  modelling the DM relative velocity distribution using a MB distribution introduces a degree of error into the $\widetilde{\cal J}_s$-factors in all three velocity-dependent models, which depends on the specific model. Nevertheless, for all three velocity-dependent models, the errors introduced in the $\widetilde{\cal J}_s$-factors due to the MB modelling of the velocity distributions are much smaller than the errors introduced due to using the under-dense empirical density profiles.

\begin{figure}
\begin{centering}
\includegraphics[width=0.98\textwidth]{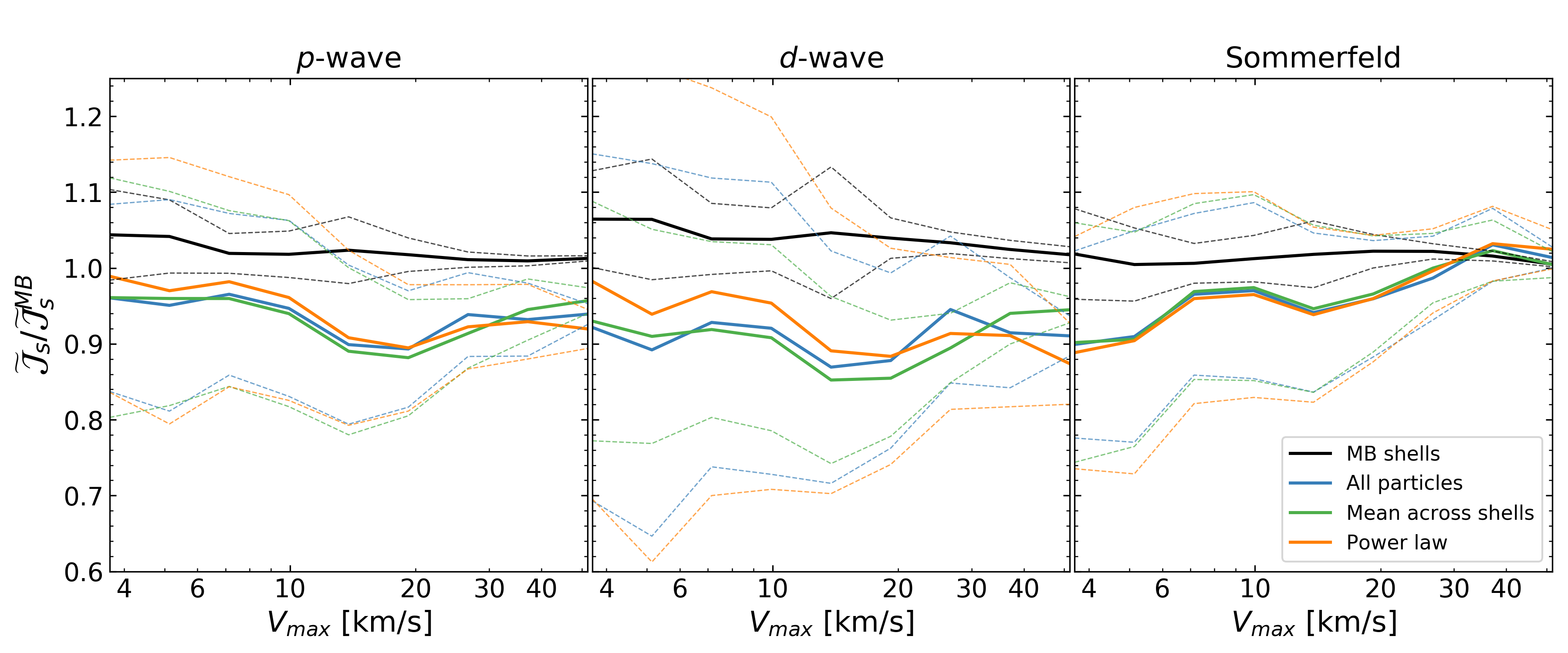}
\par\end{centering}
\caption{The ratio of the $\widetilde{\cal J}_s$-factors obtained from the empirical DM relative velocity distributions and those found from fitting a MB distribution: in different radial shells (black), to all particles in the dSph (blue), using the mean peak speed found from the best fit MB peak speeds across all shells (green), and using the peak speed found from eq.~\eqref{eq:pow-law} (orange). The solid lines show the mean ratios as a function of the dSph's maximum circular velocity, while the dashed lines show the upper and lower $1\sigma$ uncertainties. The left, middle, and right panels show the results for the p-wave, d-wave, and Sommerfeld models, respectively.}
\label{fig:corr-fac-composite}
\end{figure}

We can quantify more precisely the error introduced in the $\widetilde{\cal J}_s$-factors if we model the relative velocity distribution of the dSph as a MB distribution. For this, we compare the $\widetilde{\cal J}_s$-factor of all subhalos using the DM relative velocity distributions extracted from the simulation data with those computed using (a) the best fit MB velocity distribution in different radial shells, (b) the best fit MB distribution for all particles in the subhalo, (c) a MB distribution with a peak speed set to the mean of the best fit MB peak speeds across all radial shells, and (d) a MB distribution with a peak speed found using eq.~\eqref{eq:pow-law}. The ratios of the $\widetilde{\cal J}_s$-factors obtained directly from the simulation data  to those computed from a MB velocity distribution, $\widetilde{\cal J}_s^{\rm MB}$, using each of the above methods are shown in figure \ref{fig:corr-fac-composite} for the p-wave, d-wave, and Sommerfeld models. 

We can see from figure \ref{fig:corr-fac-composite} that fitting a MB distribution in each radial shell minimizes the error for all annihilation models. The errors introduced by the other three methods are similar to each other, especially for the p-wave and Sommerfeld models. In table \ref{tab:MB-errors}, we quantify the errors introduced in the $\widetilde{\cal J}_s$-factors using all four methods for computing the velocity distribution. At high $V_{\rm max}$, we find that fitting a MB distribution in each radial shell introduces an average error of 1.06-1.70\%, using the best fit MB distribution for all particles in a subhalo introduces an average error of 2.17-6.50\%, using the mean of MB fits across all radial shells introduces an average error of 2.45-5.91\%, and using a MB distribution from eq.~\eqref{eq:pow-law} introduces an average error of 2.93-14.86\% for the velocity-dependent annihilation models. We find that the average percent error increases at lower $V_{\rm max}$ for all annihilation models. We also find that the average percent errors are generally smaller for the Sommerfeld model in each method of calculation.

\begin{table}
\footnotesize
\makebox[\linewidth]{
\begin{tabular}{|l|c|c|c|c|}
\hline
 & MB Shells & All particles & Mean across shells & Power law \\ 
 \hline
 p-wave & [4.14, 1.25]  & [11.24, 6.50] & [12.16, 4.54]& [13.25, 8.82]\\
 \hline
 d-wave & [5.99, 1.70]  & [24.04, 9.92] & [25.30, 5.91]& [26.38, 14.86]\\
 \hline
 Sommerfeld & [2.00, 1.06]  & [12.90, 2.17] & [13.90, 2.45]& [14.43, 2.93]\\
 \hline
\end{tabular}
}
\caption{The average percent error of the $\widetilde{\cal J}$-factors of each method of calculation for the velocity-dependent annihilation models. Shown are the average percent errors for subhalos with $3.09 \leq V_{\rm max} \leq 4.30$ km/s (left numbers in the intervals) and subhalos with $43.16 \leq V_{\rm max} \leq 60.0$ km/s (right numbers in the intervals).}
\label{tab:MB-errors}
\end{table}

We note that the computational time saved using a MB distribution 
with a peak speed found from the power law fit (eq.~(3.7)) to model the relative velocity distribution rather than extracting it from the simulation data directly is substantial. In particular, we found that for our subset of subhalos selected from 
the AP-L1 simulations, the time it takes to compute the  $\widetilde{\cal J}_s$-factors using the empirical data is $\sim 20,000$ times longer than the time it takes to compute them using the MB distribution. 

\begin{figure}
\begin{centering}
\includegraphics[width=0.98\textwidth]{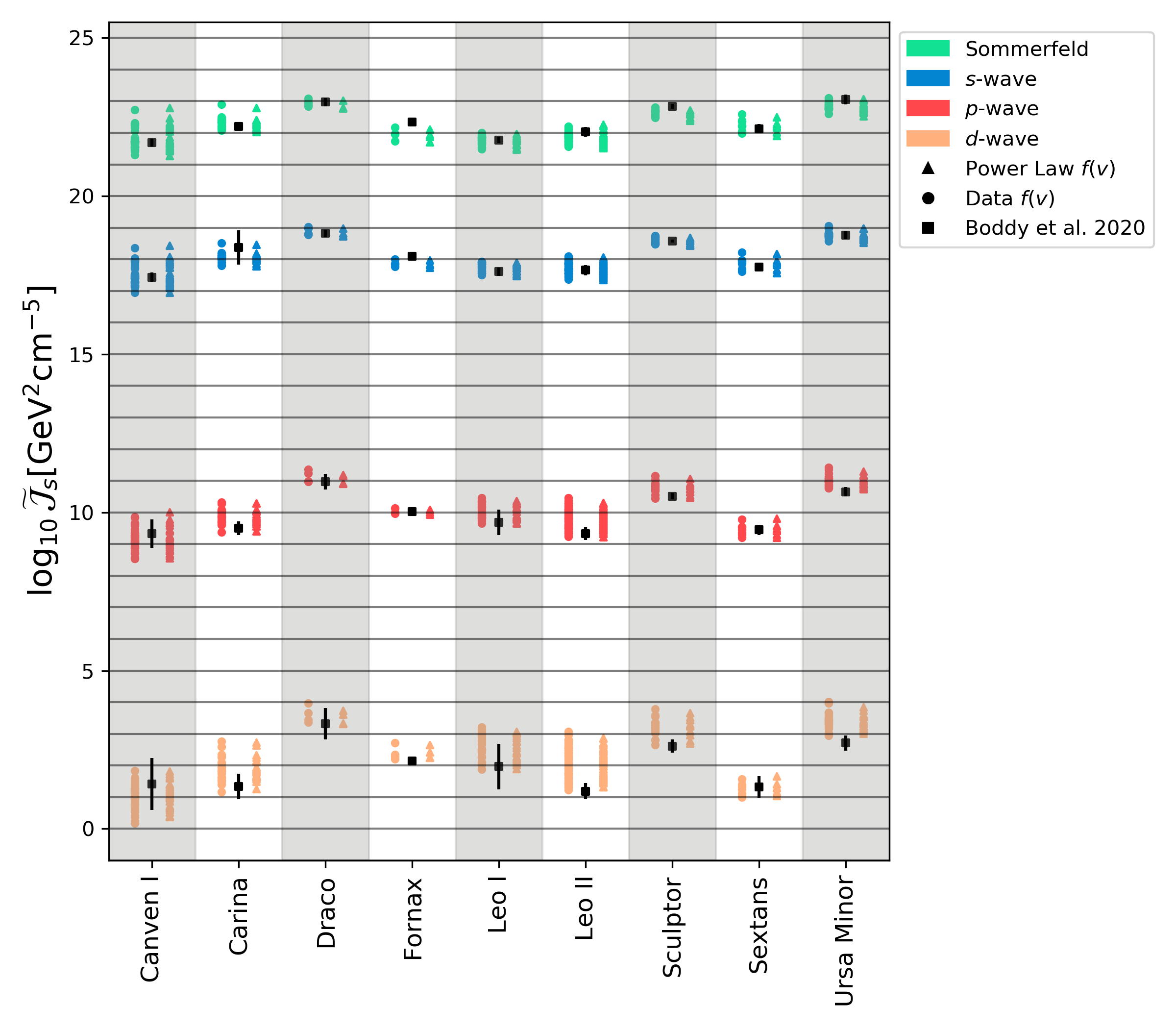}
\par\end{centering}
\caption{Integrated $\widetilde{\cal J}_s$-factors over cones with opening half-angles of $0.5^\circ$ for nine dSph analogues for the s-wave (blue), p-wave (red), d-wave (orange), and Sommerfeld (green) models. The  the $\mathcal{J}$-factors are calculated using the DM relative velocity distributions found empirically (circle), using a MB distribution with a  peak speed from the power law in eq.~\eqref{eq:pow-law} (triangle), and comparing to the data in ref.~\cite{Boddy:2019qak} (square).}
\label{fig:Boddy-compare}
\end{figure}

Next, we compare our $\widetilde{\cal J}_s$-factor results to those recently found in the literature for specific dSphs~\cite{Boddy:2019qak}. In figure \ref{fig:Boddy-compare}, we show the comparison of our $\widetilde{\cal J}_s$-factors for nine 
dSph analogues obtained using the simulation data directly and using a MB velocity distribution with a peak speed obtained from the power law fit, with those presented in figure 1 of ref.~\cite{Boddy:2019qak}. The $\widetilde{\cal J}_s$-factors in ref.~\cite{Boddy:2019qak} are calculated using eq.~\eqref{eq:Jfactor}, integrated over cones with various opening half-angles. We choose to compare to the $\widetilde{\cal J}_s$-factors integrated over a cone with an opening half-angle of $0.5^\circ$, simply because all the dSph analogues are extended out at minimum to $0.5^\circ$, but not to the the next data point of $10^\circ$. Furthermore, to compare our results to those of ref.~\cite{Boddy:2019qak}, we compute our $\widetilde{\cal J}_s$-factors by positioning the simulated dSph analogues at the same galactocentric distance as their observed dSph counterparts, obtained  from ref.~\cite{McConnachie2012}.

In ref.~\cite{Boddy:2019qak} a Navarro-Frenk-White (NFW) profile was used to model the density profiles of the dSphs, and the DM velocity distributions were assumed to be related to the density profiles by the Eddington inversion formula \cite{Widrow:2000ApJ}. This approach assumes that the DM halo is in equilibrium with a spherically symmetric potential, and that the DM has isotropic orbits. For comparison, in this work, we use the best fit Einasto density profiles for the simulated dSphs. The DM relative velocity  distributions are obtained directly from the cosmological simulations, and we also compare the results to those obtained from modeling the velocity distributions as a MB distribution. These represent the primary differences in our approach and  the approach used in ref.~\cite{Boddy:2019qak}.

As we can see in figure \ref{fig:Boddy-compare}, our results generally agree with ref.~\cite{Boddy:2019qak}. In particular, for each of the observed dSphs shown in figure \ref{fig:Boddy-compare}, the average $\widetilde{\cal J}_s$-factor over the different simulated dSph analogues is roughly on the same order of magnitude as those of ref.~\cite{Boddy:2019qak} for each annihilation model, and show no systematic trends when compared to ref.~\cite{Boddy:2019qak}. 
We also see that in general the subhalo-to-subhalo scatter is largest for d-wave and smallest for the Sommerfeld model. This trend is similar to the results obtained for the smooth halo in ref.~\cite{Board:2021bwj}.

\section{Discussion and conclusions}
\label{sec:disc}

In this paper we have presented a systematic study of the velocity-dependent dark matter annihilation signals from dwarf spheroidal galaxy analogues in the APOSTLE cosmological simulations. We extract the dark matter density and pair-wise relative velocity distributions of the simulated dwarf spheroidal analogues to compute the $\mathcal{J}$-factors in the s-wave, p-wave, d-wave, and Sommerfeld annihilation models. We also quantify the error introduced in the $\mathcal{J}$-factors if we model the dark matter relative velocity distribution as a Maxwell-Boltzmann distribution.

 We find a good agreement between the dark matter relative velocity distributions of the dwarf spheroidal analogues extracted from the simulations and their best fit Maxwellian distribution at all radii. As a result, the error introduced in the $\mathcal{J}$-factors from using the best fit Maxwellian distributions in different radial shells within a dwarf spheroidal analogue is on average 2.7\% for the three velocity-dependent models. 
 This error increases to 10.7\% 
 if we use a Maxwellian distribution with a peak speed set to the mean of the best fit Maxwellian peak speeds across all radial shells in a dwarf spheroidal analogue. If instead, we use all the dark matter particles in a dwarf spheroidal analogue to find the best fit Maxwellian distribution, the error is slightly increased to 11.1\%.  
 We also find that the latter two methods of determining the best fit peak speed of the Maxwellian distribution by using either all the dark matter particles in the dwarf spheroidal analogue or by finding the mean of the best fit peak speeds at different radii, lead to a power law relation between the best fit peak speed and the maximum circular velocity of the dwarf spheroidal analogue (i.e.~eq.~\eqref{eq:pow-law}). Using this power law to compute the $\mathcal{J}$-factors substantially reduces the computation time, and introduces on average a 13.4\% error.

We have investigated the effects of modelling the dark matter density profile of the dwarf spheroidal analogues using an Einasto profile. We find that the difference between the $\mathcal{J}$-factors calculated directly from the simulation data and those calculated from the best fit Einasto profiles is much larger than the error introduced in the calculations due to modeling the dark matter relative velocity distributions with a Maxwell-Boltzmann distribution. The large differences in the two density profiles originate from the resolution of the simulations which results in the empirical density profiles being under-dense in the inner regions of the dwarf spheroidal analogues compared to the Einasto profile.

We also find that the systematic uncertainties introduced in the integrated $\mathcal{J}$-factors by using different analogues of the same observed dwarf spheroidal in the simulations is in general larger than the error introduced by modeling the relative dark matter velocity distributions as a Maxwell-Boltzmann distribution. Additional systematic uncertainties can also be introduced by using different models for the dark matter density profiles, such as the NFW profile and its cored versions, or by breaking the assumption of spherically symmetric profiles. However, a detailed analysis of such uncertainties is beyond the scope of this work.

Finally, we compare our results to those found in the literature for specific dwarf spheroidal galaxies. We find that our integrated $\mathcal{J}$-factors are generally in good agreement with those presented in previous work, which uses simplified models for the dark matter velocity distributions of the dwarf spheroidals~\cite{Boddy:2019qak}. The simulations we use in this paper do not force strict model assumptions on the velocity distribution. 
As a result, we specifically show that the halo-to-halo scatter in the $\mathcal{J}$-factors dominate the astrophysical uncertainties, with the largest scatter for the d-wave models and the smallest for the Sommerfeld models. Such a scatter is likely to have important implications for bounds on the dark matter annihilation cross section that have been obtained in previous studies~\cite{Fermi-LAT:2016uux}.

\acknowledgments

KB and NB acknowledge the support of the Natural Sciences and Engineering Research Council of Canada (NSERC), funding reference number RGPIN-2020-07138. EP and LES acknowledge support from DOE Grant de-sc0010813. AF is supported by a UKRI Future Leaders Fellowship (grant no.~MR/T042362/1). CSF acknowledges European Research Council (ERC) Advanced Investigator 
grant DMIDAS (GA 786910). This work used the DiRAC@Durham facility managed 
by the Institute for Computational Cosmology on behalf of the STFC DiRAC 
HPC Facility (www.dirac.ac.uk). The equipment was funded by BEIS capital 
funding via STFC capital grants ST/K00042X/1, ST/P002293/1, ST/R002371/1 
and ST/S002502/1, Durham University and STFC operations grant 
ST/R000832/1.  DiRAC is part of the National e-Infrastructure.

\bibliographystyle{JHEP}
\bibliography{mainbib}

\end{document}